\documentclass[useAMS,usenatbib,twocolumn]{mn2e}

\usepackage{float,textcomp,graphicx,amsmath,pdflscape,amssymb,setspace,subfigure}

\newcommand{\Lbol}{\ifmmode L_{\rm bol} \else $L_{\rm bol}$\fi}
\newcommand{\Ledd}{\ifmmode L_{\rm edd} \else $L_{\rm edd}$\fi}
\newcommand{\Lnu}{\ifmmode L_{\rm \nu} \else $L_{\rm \nu}$\fi}
\newcommand{\Llambda}{\ifmmode L_{\rm \lambda} \else $L_{\rm \lambda}$\fi}
\newcommand{\LA}{\ifmmode L_{\rm 5100} \else $L_{\rm 5100}$\fi}
\newcommand{\Lg}{\ifmmode L_{\rm g(r)} \else $L_{\rm g(r)}$\fi}
\newcommand{\Ln}{\ifmmode L_{\rm n(r)} \else $L_{\rm n(r)}$\fi}
\newcommand{\Lo}{\ifmmode L_{\rm 0} \else $L_{\rm 0}$\fi}
\newcommand{\Ro}{\ifmmode R_{\rm 0} \else $R_{\rm 0}$\fi}
\newcommand{\Rhalf}{\ifmmode R_{\rm 1/2} \else $R_{\rm 1/2}$\fi}
\newcommand{\Rg}{\ifmmode R_{\rm g} \else $R_{\rm g}$\fi}
\newcommand{\Rms}{\ifmmode R_{\rm ms} \else $R_{\rm ms}$\fi}
\newcommand{\Rin}{\ifmmode R_{\rm in} \else $R_{\rm in}$\fi}
\newcommand{\Rout}{\ifmmode R_{\rm out} \else $R_{\rm out}$\fi}
\newcommand{\RISCO}{\ifmmode R_{\rm in} \else $R_{\rm in}$\fi}
\newcommand{\ro}{\ifmmode r_{\rm 0} \else $r_{\rm 0}$\fi}
\newcommand{\rhalf}{\ifmmode r_{\rm 1/2} \else $r_{\rm 1/2}$\fi}
\newcommand{\rg}{\ifmmode r_{\rm g} \else $r_{\rm g}$\fi}
\newcommand{\rms}{\ifmmode r_{\rm ms} \else $r_{\rm ms}$\fi}
\newcommand{\rin}{\ifmmode r_{\rm in} \else $r_{\rm in}$\fi}
\newcommand{\rout}{\ifmmode r_{\rm out} \else $r_{\rm out}$\fi}
\newcommand{\rISCO}{\ifmmode r_{\rm in} \else $r_{\rm in}$\fi}
\newcommand{\Fnu}{\ifmmode F_{\nu} \else $F_{\nu}$\fi}
\newcommand{\Flambda}{\ifmmode F_{\lambda} \else $F_{\lambda}$\fi}
\newcommand{\Mdot}{\ifmmode \dot{M} \else $\dot{M}$\fi}
\newcommand{\Mrdot}{\ifmmode \dot{M}\left(r \right) \else $\dot{M}\left(r \right)$\fi}
\newcommand{\MRdot}{\ifmmode \dot{M}\left(R \right) \else $\dot{M}\left(R \right)$\fi}
\newcommand{\Mrindot}{\ifmmode \dot{M}\left(r_{ISCO} \right) \else $\dot{M}\left(r_{ISCO} \right)$\fi}
\newcommand{\Mroutdot}{\ifmmode \dot{M}\left(r_{out} \right) \else $\dot{M}\left(r_{out} \right)$\fi}
\newcommand{\MBHdot}{\ifmmode \dot{M}_{BH} \else $\dot{M}_{BH}$\fi}
\newcommand{\MBHdotexpct}{\ifmmode \dot{M}_{BHexpected} \else $\dot{M}_{BHexpected}$\fi}
\newcommand{\Medddot}{\ifmmode \dot{M}_{edd} \else $\dot{M}_{edd}$\fi}
\newcommand{\Moutdot}{\ifmmode \dot{M}_{out} \else $\dot{M}_{out}$\fi}
\newcommand{\Mindot}{\ifmmode \dot{M}_{in} \else $\dot{M}_{in}$\fi}
\newcommand{\Mwinddot}{\ifmmode \dot{M}_{wind} \else $\dot{M}_{wind}$\fi}
\newcommand{\MBH}{\ifmmode M_{\rm BH} \else $M_{\rm BH}$\fi}
\newcommand{\Mexp}{\ifmmode M_{\rm 8} \else $M_{\rm 8}$\fi}
\newcommand{\Msun}{\ifmmode M_{\odot} \else $M_{\odot}$\fi}
\newcommand{\avisc}{\ifmmode \alpha_{visc} \else $\alpha_{visc}$\fi}

\title[The Effects of Disc Winds on the Spectrum and Black Hole Growth Rate of Active Galactic Nuclei]
{The Effects of Disc Winds on the Spectrum and Black Hole Growth Rate of Active Galactic Nuclei}

\author[Slone \& Netzer]
{Oren Slone$^1$\thanks{E-mail: shtangas@gmail.com} and Hagai Netzer$^1$\\
$^1$School of Physics and Astronomy, The Sackler Faculty of Exact Sciences, Tel-Aviv University, Tel-Aviv 69978, Israel\\}

\date{Submitted 2012}

\begin{document}

\maketitle


\setstretch{1}

\begin{abstract}
\noindent
Several properties of the standard $\alpha$-disc model for active galactic nuclei (AGN) are not entirely consistent with AGN observations.  As well as such discrepencies, observations show evidence for the existence of high mass outflow winds originating from the vicinity of the active black hole (BH).  Such winds may originate from various parts of the disc and could change the local accretion rate which should alter the emitted spectral energy distribution (SED) and affect the global disc luminosity and the BH growth rate.  The new calculations presented here show the effects of several types of winds on the observed and inferred disc properties.  Some wind profiles can have a profound effect on the observed SED and can perhaps explain the poorly understood deviations of AGN spectra from standard disc spectra.  We show a factor $\sim2$ possible error in estimating the disc luminosity and larger deviations in estimating $L/\Ledd$.  The BH growth rate computed without taking the effects of wind into account may be significantly over-estimated.  We also suggest a practical way to use the observed SED in order to make first order corrections to BH growth rate and account for the effects of disc winds.
\end{abstract}

\begin{keywords}
(galaxies:) quasars: general; galaxies: nuclei; galaxies: active; accretion, accretion discs
\end{keywords}


\section{Introduction}
\label{sec:intro}

Several properties of the spectral energy distribution (SED) of Active Galactic Nulcei (AGN) can be explained by mass accretion via a disc onto a central black hole (BH).  AGN accretion discs (AD) may vary in geometry, optical depth, accretion rate and other parameters.  The basic theory of optically thick, geometrically thin ADs (known as $\alpha$-discs) is described in \cite{SS73}.  More recent studies that include Comptonization in the disc atmosphere and improved treatment of vertical structure are described in \cite{Hubeny01} and references therein.  Observational evidence, mainly in the optical-UV, suggests that the $\alpha$-disc model spectrum is a good approximation of real AGN continua.  However, there are a number of characteristics which cannot be explained by this model.  In particular, for AGNs with $\MBH\approx10^8\Msun$ and $\Mdot/\Medddot\approx0.3$, where $\Mdot$ is the accretion rate and $\Medddot$ is the Eddington accretion rate, the model predicts a power-law, $\Fnu\propto\nu^\alpha$, with $\alpha=1/3$ throughout much of the optical-NUV continuum.  \cite{VandenBerk01} and \cite{Davis07} found that for many objects observed in the Sloan Digital Sky Survey (SDSS), the $1350-4200$\AA~ power-law is considerably softer, with $\alpha\approx-0.5$.  Observations of the FUV and EUV continua show much softer power-laws \cite[][]{Telfer01,Scott04,Shull12}.  Apart from the $\alpha$-disc continuum, the overall SED is comprised of contributions from a number of other components including emission by hot dust in the NIR part of the spectrum \cite[][]{Mor11}, a conglomeration of hundreds of broad FeII lines and Balmer continuum emission \cite[the "Small Blue Bump".][]{Grandi82,Netzer85} and a power-law x-ray source \cite[][]{Laor97}.  \cite{WardDone11} tried to combine these processes in order to model a large sample of SDSS spectra but found that for a great number of objects the underlying continuum was softer than they were able to model.  Similar conclusions were reached by \cite{Davis07}, who attempted to compare observed spectral properties of a large sample of AGNs to those predicted by their models.  A detailed overview of the variations between AGN theory and observations can be found in \cite{Koratkar99}.

AGN ADs are also probably associated with high velocity winds.  X-ray observational evidence for such winds is given in numerous publications including \cite{Pounds03a}, \cite{Reeves03}, \cite{Tombesi10} and \cite{Tombesi11} and references therein.  These studies claim to detect winds, based on Fe K-shell transitions with velocities reaching considerable fractions of the speed of light.  Broad absorption line (BAL) outflows provide more evidence for material which may be ejected from the vicinity of the BH \cite[e.g.][and references therein]{Capellupo12}.  Statistical analysis of AGNs shows a correspondence between objects exibiting BAL outflows and softened optical-UV spectra \cite[][]{Ganguly07}.  Slower winds, which probably originate further from the BH, may be detected through ion or molecular lines \cite[e.g.][and references therein]{Storchi10,Sturm11}.  \cite{King10} showed that very general considerations predict high velocity winds expected from AGNs.  All these findings suggest that winds are present in a considerable fraction of AGNs.  The origin of these winds, especially those with high velocities, could be from regions close to the BH and are possibly connected with, or ejected from, the AD itself \cite[e.g.][]{KingPounds03,ProgaKallman04,Sim08}.

Winds from AGN ADs may be caused by a number of physical processes.  Three main scenarios have been suggested: radiation driven, thermally driven or magnetically driven winds.  Radiation driven winds have been studied in numerous papers \cite[][and references therein]{Shields77,Proga00,Risaliti10}.  Such processes accelerate gas through radiation force on resonant line transitions and are able to produce outflows with velocities of the order $\sim10^4$ $km\cdot s^{-1}$ arising from radii within the inner AD.  \cite{Mckee83} and later \cite{Woods96} proposed the possibility of thermal winds caused by heating of the disc by a central x-ray source driving thermal outflows at large radii.  Many other studies of such winds also exist \cite[][and references therein]{Krolik84,Chelouche05,Everett07}.  Magnetically driven winds seem to be the best candidate for the origin of high velocity winds emitted from the inner radii of ADs.  The details of such systems have been described by many researchers \cite[][and references therein]{Blandford82,Contopoulos94,Konigl94}.  A common result is a self similar wind with a constant mass loss over each decade of the disc radius.  \cite{Blandford82} showed that for standard AGNs, only a small fraction of the disc mass is ejected by such winds.

The standard $\alpha$-disc model assumes a constant accretion rate, $\Mdot$, throughout the disc.  Significant outflows must alter the local accretion rate, in turn altering the energy flux from different radii.  This will change the emitted SED and will also affect the simple relationship between disc luminosity, $L$, and the BH growth rate, $\MBHdot$.  In this paper we investigate two major aspects of the possible effects of disc outflows on AGNs: the direct effects on the emitted SED, and the ability to estimate BH growth rate based on measurements of $L$ and the global SED.  The structure of the paper is as follows: In \S2 we give a detailed description of our numerical model which includes the effects of disc winds.  In \S3 we give the results of our numerical models and show a number of examples which demonstrate trends in the SED, in $L$ and in the BH growth rate as functions of various AGN parameters and wind profiles.  Finally, in \S4 we discuss our results and compare them to AGN observations.


\section{Calculations}
\label{sec:calculations}

\subsection{Standard $\alpha$-Disc Models}
\label{subsec:alpha_theory}

Our calculations are similar to several earlier AD calculations based on the standard $\alpha$-disc model, with the addition of GR corrections and treatment of Comptonization in the disc atmosphere \cite[][]{Teukolsky72,Laor89,Hubeny01,DavisLaor11}.  For completeness, we summarise the ingredients of the model that are needed for the new calculations.

Three parameters define the emitted SED for a standard $\alpha$-disc: BH mass ($\MBH$), accretion rate ($\Mdot$) and normalized BH spin ($a$).  Viscous torque may be parameterized via the $\alpha_{visc}$ parameter as defined in \cite{SS73}.  The standard model assumes a constant $\Mdot$, independent of radius ($R$).  The BH spin defines the innermost stable circular orbit (ISCO) around the central BH and sets the efficiency, $\eta$, of the conversion of rest energy to electromagnetic radiation ($L=\eta\Mdot c^2$).  Angular momentum within the disc is transferred outward causing mass accretion onto the central BH and viscous dissipation leads to heating of the gas and emission of radiation.  A local balance is attained between work done by torque forces between adjacent differential annuli, gravitational potential energy, kinetic energy and emitted radiation.  The torque forces, which are powered by gravitational energy, may correspond to either positive or negative work done by the torque.  For the standard $\alpha$-disc, the work done by torque changes sign at a radius $R=9/4$ $\Rg$, where $\Rg=G\MBH/c^2$ is the gravitational radius.  Above this radius, energy is contributed to the emitted radiation and below this radius, energy is removed.

\begin{figure*}
\center
\includegraphics[width=0.7\textwidth]{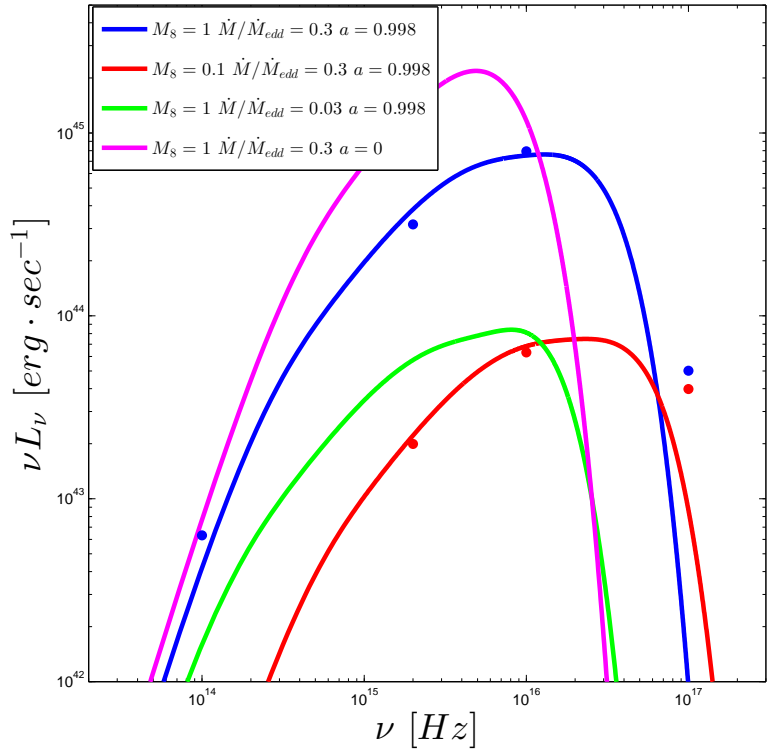}
\caption{Calculated SEDs for $\alpha$-discs with various values of $\Mexp$, $\Mdot/\Medddot$ and $a$ and comparison to earlier models.  Blue and red points correspond to values from Hubeny et al. (2001) for the same parameters as used in our models (blue and red lines respectively).  They fit well with our models for frequencies below $\sim3\times10^{16}$ $Hz$ and are harder than our models for higher frequencies (see text).\label{fig:standard_models}}
\end{figure*}

A derivation of the emissivity associated with viscous dissipation may be done using the Navier-Stokes equations for a non-compressable viscous fluid \cite[see][]{Clarke03}.  We derive the same result as theirs using the approach of \cite{King02}.  The emmisivity of the disc at various radii depends on the local gravitational energy and work done by the local torque.  We show below that both these components are altered when a wind is taken into account.  The total emmisivity is
\begin{equation}
 D(R)=D_N(R)+D_G(R),
\label{eq:DR}
\end{equation}
where
\begin{equation}
 D_G(R)=\Mdot\frac{G\MBH}{8\pi R^3}
\label{eq:DG}
\end{equation}
is the emmisivity component due to loss of gravitational and kinetic energy.  $D_N(R)$ is obtained by considering the luminosity due to the work done by the torque,
\begin{equation}
 L_N(R)=-\left(\frac{G\MBH}{R^3}\right)^{1/2} \int_{R_{in}}^{R}\Mdot\left(G\MBH\right)^{1/2}d\left(R^{1/2}\right),
\label{eq:LN}
\end{equation}
where the integrand is the differential of the local torque.  The emmisivity associated with this component is then
\begin{equation}
 D_N(R)=\frac{dL_N}{4\pi RdR}\mbox{ .}
\label{eq:DN}
\end{equation}
The effective temperature of the disc as a function of radius is
\begin{equation}
 T(r)=\left(\sigma^{-1}D(r)\right)^{1/4}=\left(\frac{3c^6}{8G^2\pi\sigma}\right)^{1/4}\left(\frac{\Mdot^{1/4}}{\MBH^{1/2}}\right)f(r)r^{-3/4},
 \label{eq:Tr}
\end{equation}
where $r=R/R_g$ and $f(r)$ is a function of order unity (for large values of $r$) which depends on the GR terms.  Assuming local blackbody (BB) emission, the isotropic $\Lnu$ is
\begin{equation}
 \Lnu=\int_{R_{in}}^{R_{out}}4\pi R\left(\pi B_\nu\left(T\right)\right)dR=\frac{8\pi^2h\nu^3}{c^2}\int_{R_{in}}^{R_{out}}\frac{RdR}{exp(x)-1},
 \label{eq:Lnu}
\end{equation}
where $x=h\nu/kT$ and $R_{in}$ and $R_{out}$ are the innermost and outermost radii of the disc respectively.  The dependence of $\Lnu$ on temperature and disc geometry is isolated in the integral
\begin{equation}
 \theta(\nu)\equiv\int_{R_{in}}^{R_{out}}\frac{RdR}{exp(x)-1}\mbox{ .}
 \label{eq:theta}
\end{equation}

Our AD calculations include Comptonization of the emitted radiation in the AD atmosphere.  We model Comptonization in the manner applied by \cite{Laor89}, namely, we calculate the parameter $y$ (their eqn. 29) which is a function of $r$ and $\nu$ and is dependant on the viscosity parameter, $\avisc$, and on the gas composition.  We have compared our values of $y$ to those of \cite{Hubeny01} and find them to be in good agreement.

Fig. \ref{fig:standard_models} shows some examples of our standard disc calculations and a comparison with those calculated by \cite{Hubeny01}.  The agreement is good, except for very high frequencies where our SEDs are somewhat softer, mainly because we do not account of radiative transfer within the disc atmosphere.  Fig. \ref{fig:standard_models} also demonstrates the known dependence of the SED shape on $\Mexp$ (defined by $\Mexp\equiv\MBH/(10^8\Msun)$), $\Mdot/\Medddot$ and $a$.  BH spin governs the radius of the ISCO, changing the spectral shape through the boundaries of the integral $\theta(\nu)$ (eqn. \ref{eq:theta}).  $\MBH$ and $\Mdot$ change the spectral shape through $T(r)$.

The $\alpha$-disc model leads to a simple relation between the luminosity and $\Mdot$,
\begin{equation}
\frac{L}{\Ledd}=\frac{\Mdot}{\Medddot},
\label{eq:L=Mdot}
\end{equation}
where $L=\int_0^\infty\Lnu d\nu$, $\Ledd$ scales linearly with $\MBH$ and $\Medddot$ depends on $\MBH$ and $a$.  As shown below, this simple relation breaks down when $\Mdot$ is no longer constant.

\subsection{Wind Ejecting $\alpha$-Disc Models}
\label{subsec:model}

\begin{figure*}
\center
\includegraphics[width=0.7\textwidth]{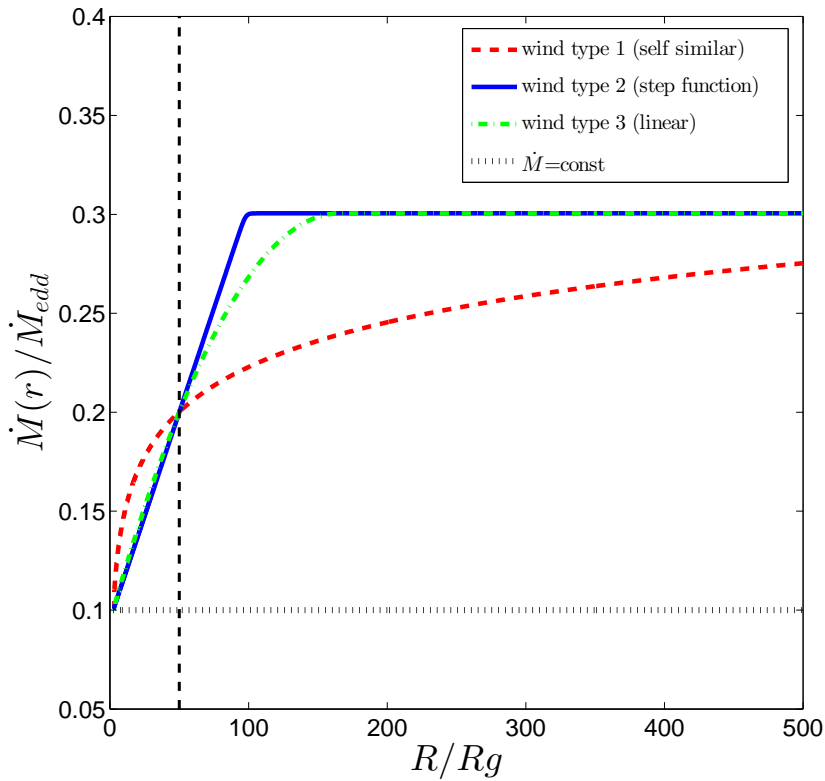}
\caption{Example $\Mrdot$ profiles for models with $\Mexp=1$, $\Mindot/\Medddot=0.1$, $\Moutdot/\Medddot=0.3$ and $a=0.9$.  $\rhalf$ is the radius at which $50\%$ of total wind mass has been ejected from the disc.  All models shown here have  $\rhalf=50$ (vertical dashed line).  The dotted black line is a standard $\alpha$-disc with constant $\Mdot/\Medddot=0.1$.\label{fig:Mrdot_a=09}}
\end{figure*}

We now explore how winds ejected directly from the surface of the disc can affect the SED.  Ejection of mass from various radii causes $\Mdot$ to change as a function of $r$, thus $\Mdot$ is replaced by $\Mrdot$.  We define $\Medddot$ exactly as above, thus, it's value is independent of the existence of wind.  We assume the local $\Mrdot$ to be approximately constant within each differential annulus, allowing us to calculate the local emmisivity of each annulus in a way similar to the standard $\alpha$-disc model.  We assume a certain mass inflow rate, $\Mindot$, at the ISCO.  We then specify a radius dependent mass outflow rate (hereafter ``wind'') which defines $\Mrdot$.  We experimented with three functional forms for the wind profile: (1) a self similar wind function, such as that described by \cite{Blandford82}, (2) a step function wind profile and (3) a wind function which decreases linearly with radius.

\begin{figure*}
\center
\includegraphics[width=0.7\textwidth]{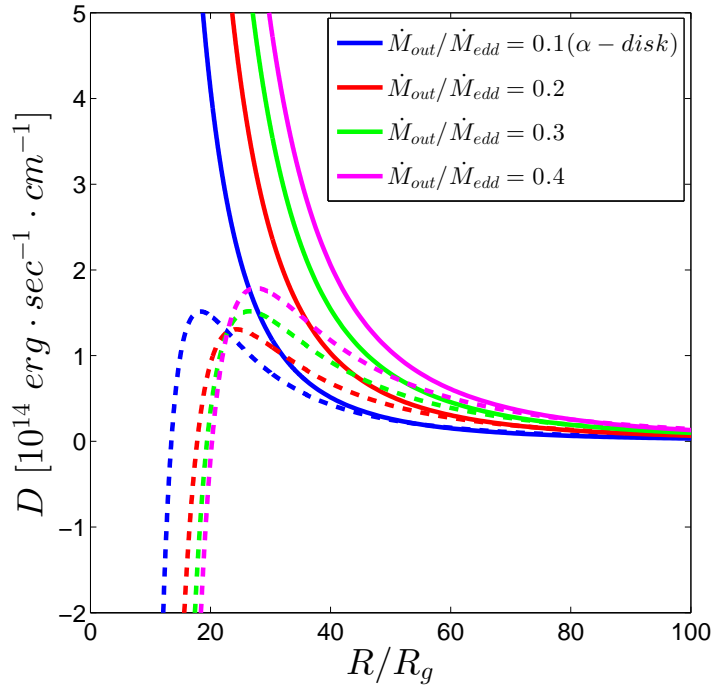}
\caption{Plots of emmisivity for various type (2) wind functions for models without Comptonization and GR effects.  All models have $\Mexp=1$, $\Mindot/\Medddot=0.1$, $a=0$, $\avisc=0.1$, $b_2=0.4$ and $\rhalf=10$.  Blue lines correspond to a standard $\alpha$-disc model with constant $\Mdot$.  Solid lines correspond to emissivity associated with gravitational and kinetic energy loss, $D_G(r)$.  Dashed lines correspond to emissivity associated with work done by torque, $D_N(r)$.  $D_G(r)$ is modified locally proportionally to $\Mrdot$ at each radius.  Regarding $D_N(r)$, the wind profile changes $\Mrdot$ such that the integral in eqn. \ref{eq:LN} is modified, thereby changing the form of $D_N(r)$.  As more mass is lost throughout the disc, torque forces do negative work and consume energy over a larger range of radii (between $\rin$ and the radius where $D_N(r)=0$). \label{fig:Dg_Dn}}
\end{figure*}

We define $\rhalf$ to be the radius at which $50\%$ of total wind mass has been ejected from the disc (see fig. \ref{fig:Mrdot_a=09}).  The self similar, type (1) wind function, is of the form
\begin{equation}
 \frac{d\Mwinddot(r)}{dr}=\frac{a_1}{r},
\label{eq:wind1}
\end{equation}
where $a_1$ is a normalization parameter and depends on $\Mindot$ and $\Moutdot$.  For such a wind and for $\Mexp=1$, $a=0.9$ and a disc with a maximum radius of $1080$ $R_g$, we get $\rhalf=50$.
The type (2) wind profile resembles a step function and depends on two parameters, $b_1$ and $b_2$,
\begin{equation}
 \frac{d\Mwinddot(r)}{dr}=b_1\left(1-\left[1+\exp\left(-2b_2\left(r-r_m\right)\right)\right]^{-1}\right),
\label{eq:wind2}
\end{equation}
where $r_m$ is a parameter which defines $\rhalf$.  The type (3) function is a linear function of the form
\begin{equation}
 \frac{d\Mwinddot(r)}{dr}=(c_1-c_2\cdot r),
\label{eq:wind3}
\end{equation}
where $c_1$ and $c_2$ are chosen such that $\rhalf$ has the same definition as above.  In all three cases, the wind function defines $\Mrdot$ according to
\begin{equation}
 \Mrdot=\Mindot+\int_{r_{in}}^{r}d\Mwinddot(r).
\label{eq:Mdot_Mwinddot}
\end{equation}
Diagrams of $\Mrdot$ for some of the examples discussed in this paper are shown in fig. \ref{fig:Mrdot_a=09}.  We found very similar results for winds of types (2) and (3) (see details in \S\ref{sec:results}), and significant differences for winds of type (1).  In this work we show results for wind types (1) and (2).

The varying $\Mrdot$ has a number of effects on the disc geometry and the emitted SED\footnote{Note that we do not consider the fate of the wind after it leaves the disc.}.  The removal of energy from the AD changes the balance between the gravitational potential and torque energies within the disc.  $D_G(r)$, $D_N(r)$ and $T(r)$ may be calculated locally according to equations \ref{eq:DG}, \ref{eq:LN}, \ref{eq:DN} and \ref{eq:Tr}, with the values of $\Mrdot$ at each annulus modified according to the wind function.  The gravitational component, $D_G(r)$, is simply proportional to $\Mrdot$.  The dependence of the torque component on $\Mrdot$ is less straightforward.  The radius where $D_N(r)=0$ is no longer $r=9/4$.  Fig. \ref{fig:Dg_Dn} shows $D_N(r)$ and $D_G(r)$ for four different values of $\Moutdot$ and the same $\Mindot$, calculated without including the effects of Comptonization and GR (discussed below).  The modifications of the emissivity change  $T(r)$, and accordingly $\Lnu$ is altered through the modification of the function $\theta(\nu)$ (eqn. \ref{eq:theta}).

To the best of our knowledge, GR corrections, which have the effect of cooling the inner radii of the disc, have never been calculated for a wind ejecting $\alpha$-disc.  We have therefore adopted the following procedure.  We calculated the function
\begin{equation}
 F(r)=\frac{T_{rel}(r)}{T_{newt}(r)}
\label{eq:TrelNewt}
\end{equation}
using the standard $\alpha$-disc model (i.e. $\Mdot=const$), where $T_{Newt}(r)$ and $T_{rel}(r)$ are Newtonian and GR temperatures respectively.  The values of the function $F(r)$ depend only on the gravitional potential and therefore only on $a$.  For the wind calculations we obtain $T(r)$ numerically and multiply by $F(r)$.  This has the effect of cooling the inner radii of our models, where relitavistic effects are most prominent.

In wind ejecting discs, the BH growth rate no longer depends on the accretion rate at the outer radii of the disc, only on $\Mindot$.  Thus, eqn. \ref{eq:L=Mdot} is no longer valid and measurement of $L$ is no longer a direct measure of $\MBHdot$ even if $a$ is known.  We later discuss the implications and possible errors in estimating BH growth rates when the effects of winds are ignored.

\begin{table*}
\center
\caption{Parameters of the models shown in this paper.  All models have $\Mexp=1$ and $\alpha_{visc}=0.1$.  Models with wind type (2) (eqn. \ref{eq:wind2}) have $b_2=0.4$.  Models 1.1, 2.1, 1.5 and 2.5 are standard $\alpha$-discs with constant $\Mdot$.  All other models are wind ejecting $\alpha$-discs.  Models 1.2 - 1.8 and 2.2 - 2.8 were calculated with wind type (2), models 3.3, 3.6, 3.7 and 3.8 were calculated with wind type (1) and have the same parameters as models 2.3, 2.6, 2.7 and 2.8 respectively.  Models with wind type (2) are divided into categories with either the same $\Mindot$ and $\Moutdot$ and differing $\rhalf$ (category A) or the same $\Mindot$ and $\rhalf$ and differing $\Moutdot$ (category B).  Category (C) models are 1.3, 1.8, 2.3, 2.8, 3.3 and 3.8 which have the same $\Moutdot$ and $\rhalf$ and differing $\Mindot$.}
\begin{tabular}{l|c|c|c|c|c|c}
\hline
\hline
Model       & BH spin (a) & \Mindot    & \Moutdot   & \rhalf & Wind Type & Category \\
            &             & [\Medddot] & [\Medddot] &        &           &          \\
\hline
\hline

Model 1.1   & 0           & 0.1        & 0.1        & -      & -         &          \\

Model 1.2   & 0           & 0.1        & 0.3        & 10     & 2         & A        \\

Model 1.3   & 0           & 0.1        & 0.3        & 50     & 2         &          \\

Model 1.4   & 0           & 0.1        & 0.3        & 250    & 2         &          \\

\hline

Model 1.5   & 0           & 0.01       & 0.01       & -      & -         &          \\

Model 1.6   & 0           & 0.01       & 0.03       & 50     & 2         & B        \\

Model 1.7   & 0           & 0.01       & 0.1        & 50     & 2         &          \\

Model 1.8   & 0           & 0.01       & 0.3        & 50     & 2         &          \\

\hline
\hline

Model 2.1   & 0.9         & 0.1        & 0.1        & -      & -         &          \\

Model 2.2   & 0.9         & 0.1        & 0.3        & 10     & 2         & A        \\

Model 2.3   & 0.9         & 0.1        & 0.3        & 50     & 2         &          \\

Model 2.4   & 0.9         & 0.1        & 0.3        & 250    & 2         &          \\

\hline

Model 2.5   & 0.9         & 0.01       & 0.01       & -      & -         &          \\

Model 2.6   & 0.9         & 0.01       & 0.03       & 50     & 2         & B        \\

Model 2.7   & 0.9         & 0.01       & 0.1        & 50     & 2         &          \\

Model 2.8   & 0.9         & 0.01       & 0.3        & 50     & 2         &          \\

\hline

Model 3.3   & 0.9         & 0.1        & 0.3        & 50     & 1         &          \\

Model 3.6   & 0.9         & 0.01       & 0.03       & 50     & 1         &          \\

Model 3.7   & 0.9         & 0.01       & 0.1        & 50     & 1         &          \\

Model 3.8   & 0.9         & 0.01       & 0.3        & 50     & 1         &          \\

\hline
\hline
\end{tabular}
\label{table:models_shown}
\end{table*}

We tested our model for a large range of $\MBH$, a, $\Mrdot$ and $\alpha_{visc}$.  $\Mrdot$ is characterized by $\Mindot$, $\Moutdot$, $\rhalf$ and the wind function.  We show here only the most interesting results and only those with wind function types (1) and (2) (eqns. \ref{eq:wind1} and \ref{eq:wind2}).  The results are almost independent of our choice of $\alpha_{visc}$ which we have chosen to be $0.1$ in all our calculations.  In table \ref{table:models_shown} we give details of the models shown in this paper.  The models have all been calculated with $\Mexp=1$ and each with differing BH spin, $\Mindot$, $\Moutdot$ and values of $\rhalf$.  It is convenient to compare different models by deviding them into three catagories: (A) keeping $\Mindot$ and $\Moutdot$ constant while differing $\rhalf$, (B) keeping $\Mindot$ and $\rhalf$ constant while differing $\Moutdot$ and (C) keeping $\Moutdot$ and $\rhalf$ constant while differing $\Mindot$.  Models in table \ref{table:models_shown} are classified according to BH spin, category (A) or (B), and wind type (1) or (2).  Comparing models 1.3 and 1.8, models 2.3 and 2.8 and models 3.3 and 3.8 gives category (C).  For models with wind type (1), the radius of the outer edge of the disc was chosen such that $\rhalf=50$.  Note that models with the same $\Mindot/\Medddot$ and different $a$ (e.g. models 1.1 and 2.1) do not have the same accretion rate, $\Mdot$, because $\Medddot$ depends on spin.  Each model was used to calculate a number of parameters which define the shape of the SED, the total luminosity, $L$, and the BH growth rate, $\MBHdot=(1-\eta)\Mindot$.

Energy conservation requires that the total loss of gravitational and kinetic energy within the disc is equal to the total emitted luminosity. Our calculations do not consider the fate of the mass within the wind after it has left the disk and hence it only includes half the energy required for the wind mass to escape the BH potential.  There are a number of possibilities regarding the fate of this mass which are consistent with our model.  One such possibility is that the ejected wind rises above the surface of the AD, but does not leave the physical vicinity of the BH and therefore does not require additional kinetic energy.  This wind may re-enter the disc at larger radii or may accumulate and become ionized in regions above the disc (creating a ``failed wind'').  Another possibility may be that the wind is not isotropic (as suggested by many models) and that it escapes the system by absorbing radiation propogating at angles closely parallel to the disc.  Therefore, if we view the AD at angles close to face on, the radiation we observe may not be influenced by this effect.  Of course, there are other possibilities that involve additional acceleration of the wind by part of the radiation that we included in our luminosity budget.  A detailed discussion of these and similar possibilities is beyond the scope of our paper.


\section{Results}
\label{sec:results}

In the following section we analyze each model's SED and compare the spectra of discs with various parameters to one another.  Two major aspects of the wind ejecting disc models are discussed: the spectral changes, and the effect of mass outflow on the ability to evaluate $\MBHdot$.  Results of the spectral changes are expressed by showing the approximated power-law behaviour, $\Fnu\propto\nu^\alpha$, within various wavelength ranges and by the mean weighted ionizing photon energy of the Lyman continuum, $<E_{ion phot}>$.  The degree of inaccuracy in evaluation of $\MBHdot$ from the observations depends on our understanding of the effect of winds on the SED and on the ratio of the measured $L$ to the luminosity that would have been observed from a standard (constant $\Mdot$) $\alpha$-disc with $\Mdot=\Mindot$.  We denote this luminosity $L_0$, which for a wind ejecting model satisfies the equation
\begin{equation}
 \frac{\Lo}{\Ledd}=\frac{\Mindot}{\Medddot}\mbox{ }.
\label{eq:Lo=Mdot}
\end{equation}

\begin{table*}
\center
\caption{Continuum power-laws ($\Lnu\propto\nu^{\alpha}$) and mean ionizing photon energies for models shown in table \ref{table:models_shown}.}
\begin{tabular}{l|c|c|c|c|c}
\hline
\hline
Model       & $\alpha_{456-912\AA}$ & $\alpha_{912-1450\AA}$ & $\alpha_{1450-4200\AA}$ & $\alpha_{4200-5100\AA}$ & $<E_{ion phot}>$ \\
            &                       &                        &                         &                         & $[eV]$         \\
\hline
\hline

Model 1.1   & -1.52                 & -0.53                  & -0.044                  & 0.19                    & 21.32          \\

Model 1.2   & -1.31                 & -0.43                  & 0.016                   & 0.33                    & 22.07          \\

Model 1.3   & -1.66                 & -0.75                  & -0.21                   & 0.20                    & 21.10          \\

Model 1.4   & -1.55                 & -0.60                  & -0.20                   & 0.063                   & 21.27          \\

\hline

Model 1.5   & -3.58                 & -1.47                  & -0.40                   & -0.0053                 & 17.86          \\

Model 1.6   & -3.62                 & -1.61                  & -0.61                   & -0.17                   & 17.83          \\

Model 1.7   & -3.65                 & -1.80                  & -0.71                   & -0.15                   & 17.81          \\

Model 1.8   & -3.41                 & -1.66                  & -0.56                   & 0.060                   & 18.06          \\

\hline
\hline

Model 2.1   & -0.61                 & -0.14                  & 0.097                   & 0.21                    & 26.47          \\

Model 2.2   & -0.79                 & -0.26                  & 0.038                   & 0.22                    & 25.53          \\

Model 2.3   & -0.72                 & -0.35                  & -0.13                   & 0.085                   & 23.098         \\

Model 2.4   & -0.63                 & -0.19                  & -0.031                  & 0.025                   & 26.40          \\

\hline

Model 2.5   & -1.59                 & -0.58                  & -0.086                  & 0.11                    & 21.09          \\

Model 2.6   & -1.64                 & -0.70                  & -0.31                   & -0.12                   & 21.01          \\

Model 2.7   & -1.80                 & -1.00                  & -0.57                   & -0.23                   & 20.78          \\

Model 2.8   & -2.03                 & -1.26                  & -0.61                   & -0.13                   & 20.44          \\

\hline

Model 3.3   & -0.69                 & -0.24                  & -0.012                  & 0.12                    & 26.14          \\

Model 3.6   & -1.63                 & -0.67                  & -0.19                   & -0.0030                 & 21.068         \\

Model 3.7   & -1.61                 & -0.72                  & -0.26                   & -0.064                  & 21.24          \\

Model 3.8   & -1.41                 & -0.65                  & -0.24                   & -0.022                  & 22.041         \\

\hline
\hline
\end{tabular}
\label{table:slopes}
\end{table*}

\subsection{Effects of Wind on the Spectral Energy Distribution}
\label{subsec:SED}

\begin{figure*}
\begin{minipage}{\textwidth}
\center
\begin{tabular}{cc}
\includegraphics[width=0.5\textwidth]{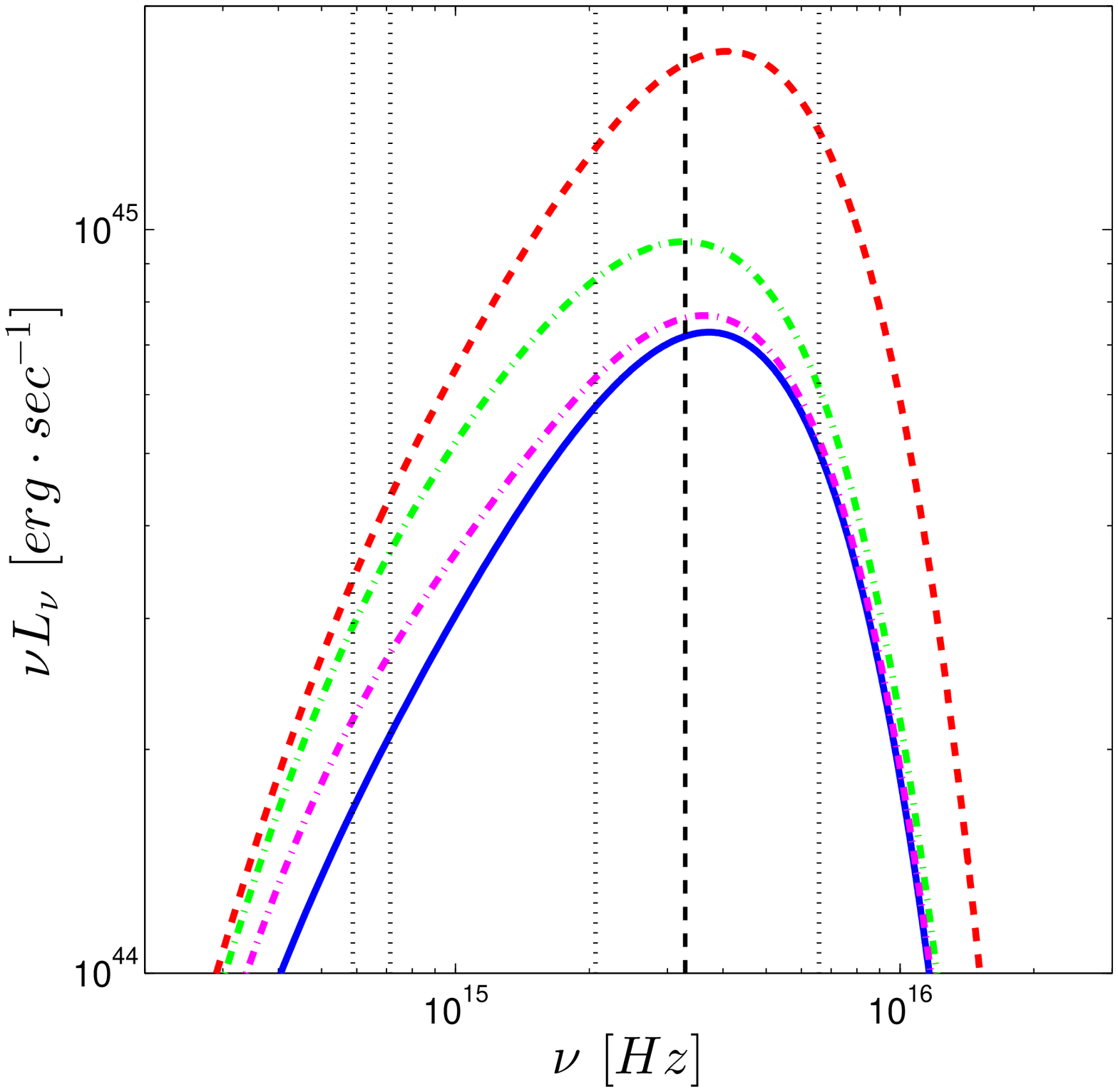} & \includegraphics[width=0.5\textwidth]{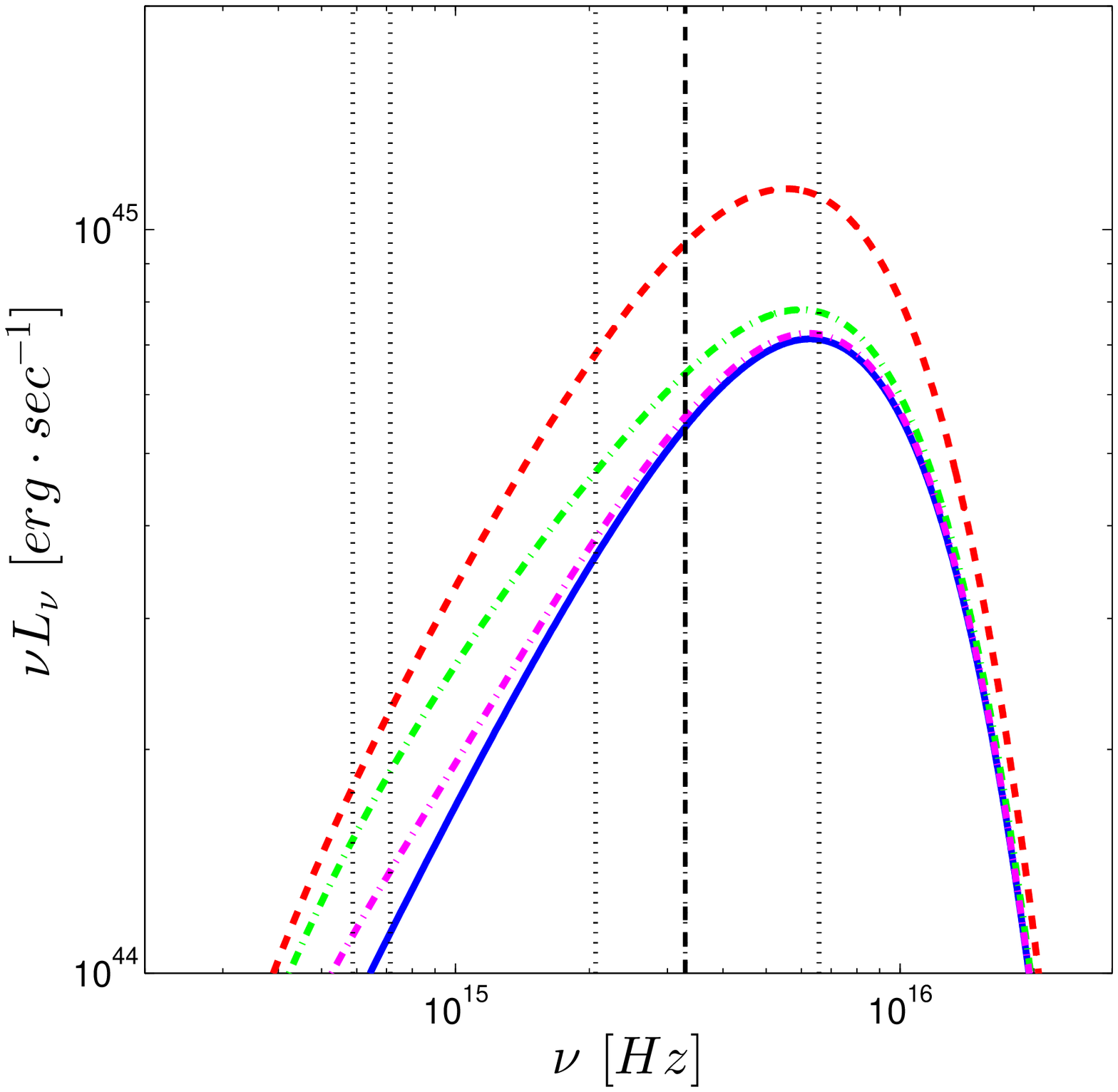}
\end{tabular}
\caption{Comparison of SEDs for models of category (A).  All have the same $\Mindot=0.1$ and therefore the same $\MBHdot$.  {\it Left}: SED comparison for models 1.1 - 1.4 with $a=0$, {\it Right}: SED comparison for models 2.1 - 2.4 with $a=0.9$.  Solid blue lines - models 1.1 and 2.1; dashed red lines - models 1.2 and 2.2; dotted dashed green lines - models 1.3 and 2.3; dotted dashed magenta lines - models 1.4 and 2.4.  Vertical dotted lines mark boundary wavelengths used for calculations of power-laws in table \ref{table:slopes}.\label{fig:cat1_specs}}
\end{minipage}
\end{figure*}

\begin{figure*}
\begin{minipage}{\textwidth}
\center
\begin{tabular}{cc}
\includegraphics[width=0.5\textwidth]{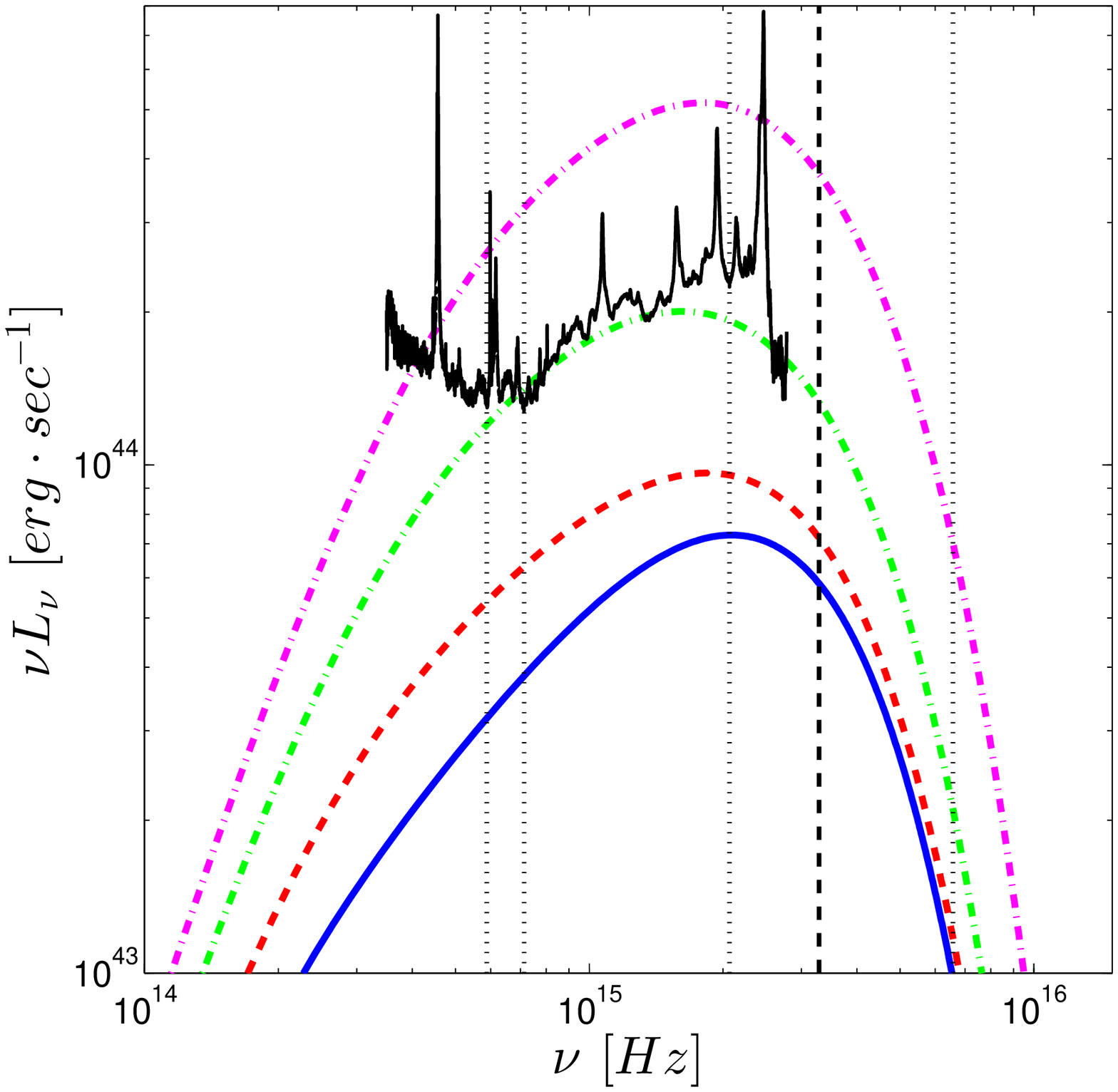} & \includegraphics[width=0.5\textwidth]{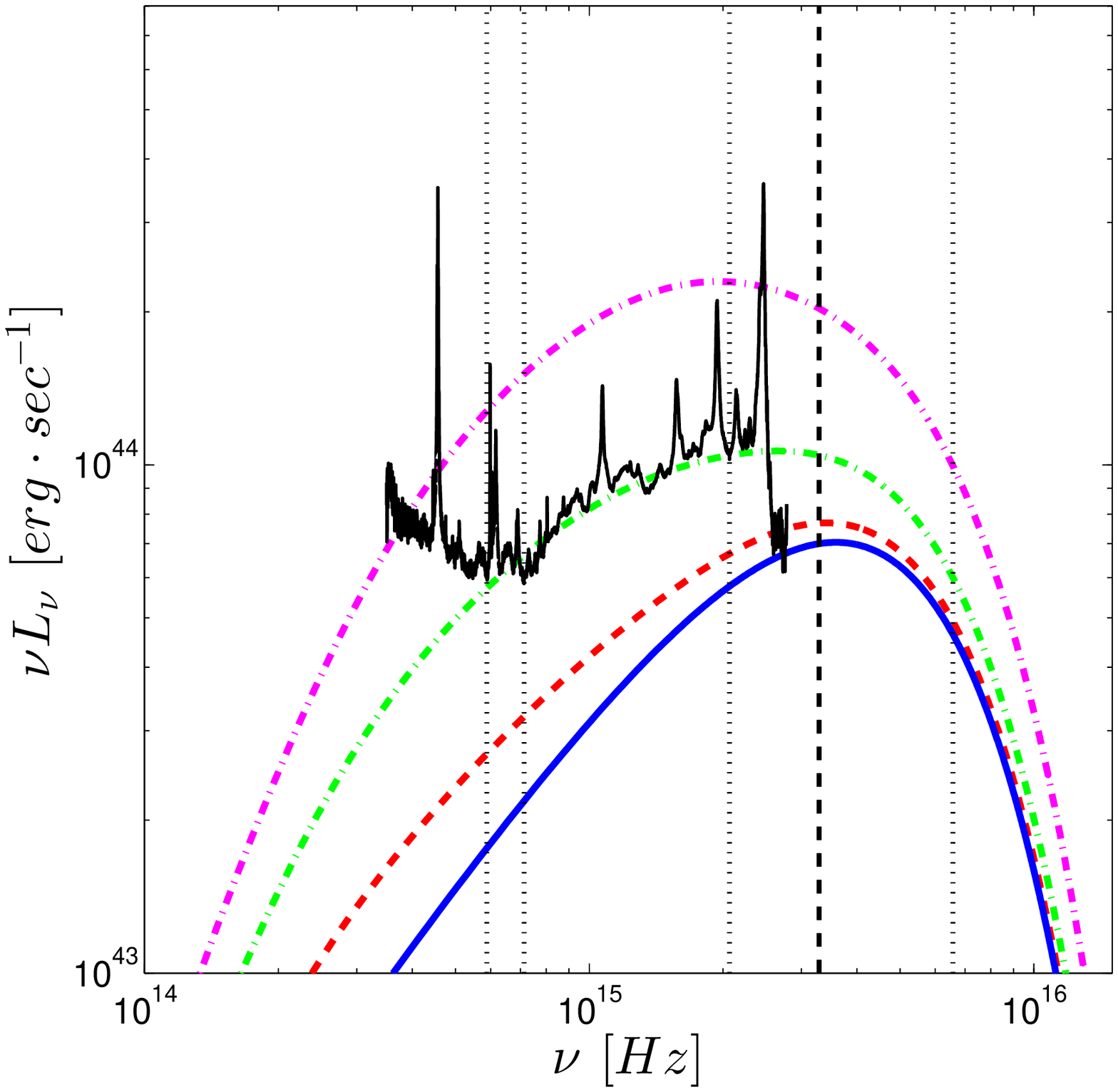}
\end{tabular}
\caption{Comparison of SEDs for models of category (B).  All have the same $\Mindot=0.01$ and therefore the same $\MBHdot$.  {\it Left}: SED comparison for models 1.5 - 1.8 with $a=0$, {\it Right}: SED comparison for models 2.5 - 2.8 with $a=0.9$.  Solid blue lines - models 1.5 and 2.5; dashed red lines - models 1.6 and 2.6; dotted dashed green lines - models 1.7 and 2.7; dotted dashed magenta lines - models 1.8 and 2.8.  Vertical dotted lines mark boundary wavelengths used for calculations of power-laws in table \ref{table:slopes}.  Also shown in black is a composite SED from Vanden Berk (2001) with arbitrary normalization.  Note that the reddening of the IR-optical continuum of the composite is caused by a significant contribution of the host galaxy.\label{fig:cat2_specs}}
\end{minipage}
\end{figure*}

We examine the SED shape by fitting a power-law, $\Fnu\propto\nu^\alpha$, to four wavelength bands: $456-912$\AA, $912-1450$\AA, $1450-4200$\AA~ and $4200-5100$\AA.  For each model we calculate these four values of $\alpha$ and the mean ionizing photon energy for photons with energies above 13.6eV.  SEDs for several models are plotted in figs. \ref{fig:cat1_specs} and \ref{fig:cat2_specs}.  The results for $\alpha$ and the mean ionizing photon energies are listed in table \ref{table:slopes}.

For a wind ejecting disc with given BH mass and spin, the spectral shape is influenced by $\Mrdot$ which can be characterized by the values of $\Mindot$, $\Moutdot$ and $\rhalf$.  In general, a large accretion rate throughout much of the disc has the effect of causing higher total luminosity and shifting the peak luminosity toward higher frequencies.  The closer the high accretion rate reaches to $\rISCO$, i.e. the lower the value of $\rhalf$, the more high energy UV radiation will be emitted.  The low energy, optical-NIR radiation, is emitted mainly from large radii and depends less on $\rhalf$.

The SEDs of the category (A) models, 1.1 - 1.4 and 2.1 - 2.4, are plotted in fig. \ref{fig:cat1_specs}.  All have $\Mindot/\Medddot=0.1$ and $\Moutdot/\Medddot=0.3$ (except for standard $\alpha$-discs 1.1 and 2.1 which have constant accretion rates).  The shapes of the SEDs of wind ejecting models as compared to standard models 1.1 and 2.1 are influenced by two competing effects: higher total luminosity together with a shift of the peak luminosity bluewards, and a sharper fall in $\Lnu$ at high frequencies, caused by reduction in accretion rate at small radii.  These two effects tend to cancel each other out in their influence on the shape of the UV spectrum as can be seen by the similar values of mean ionizing photon energies in the last column of table \ref{table:slopes}.  Comparing spectral shapes of models with wind to those of models without wind we see that the same two effects create very similar spectral shapes between models 1.1 and 1.4 and between models 2.1 and 2.4.  As $\rhalf$ grows, the spectral shape changes most in the NIR-optical continuum with the UV continuum less influenced because of less accretion at small radii.

Models compared from category (B) are 1.5 - 1.8 and 2.5 - 2.8, all with $\Mindot/\Medddot=0.01$ and $\rhalf=50$ (except for standard $\alpha$-discs 1.5 and 2.5 for which $\rhalf$ has no meaning).  Their SEDs are plotted in fig. \ref{fig:cat2_specs}.  The growing $\Moutdot$ has it's greatest effect on amplification of the NIR-optical continuum and a lesser effect on the UV continuum.  Spectral changes between models with wind to those without wind are more prominent for BHs with $a=0.9$ than for BHs with $a=0$ (as can be seen, for example, by comparing model 1.5 to 1.8 and model 2.5 to 2.8).  This is because for higher spin values, smaller radii are influenced by the wind.  The changes are most prominent at lower frequencies.

Comparison of models from category (C) (models 1.3, 1.8, 2.3, 2.8, 3.3 and 3.8) have different behaviour.  For such AGNs, with the same $\Moutdot$ and $\rhalf$ and different values of $\Mindot$, the most prominent spectral changes are in the UV spectrum.  The reason for this is the very different accretion rates at small radii and the very similar accretion rates at large radii.  We will later show that this characteristic of wind ejecting discs enables us to best constrain BH growth rate for AGNs with known luminosities.

\begin{figure*}
\center
\includegraphics[width=0.7\textwidth]{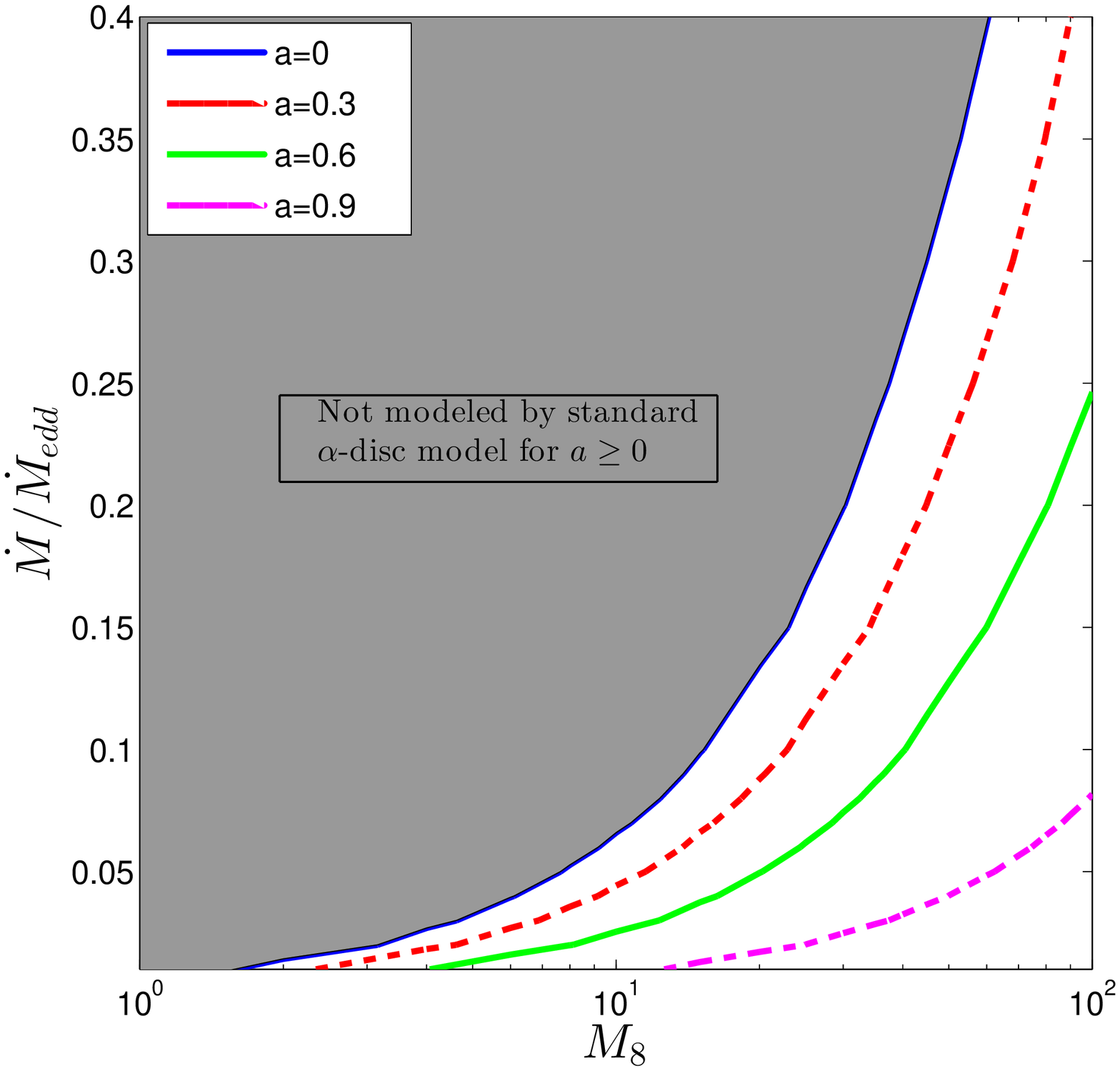}
\caption{Contours of $\alpha_{1450-4200\AA}=-0.5$ for standard $\alpha$-discs with varying $\Mexp$ and $\Mdot/\Medddot$ are shown here.  Each contour corresponds to a different value of BH spin, $a$.  The grey area corresponds to the parameter space where the standard $\alpha$-disc cannot model $\alpha_{1450-4200\AA}=-0.5$ for $a\geq0$.  Low accretion rates and large BH masses soften the optical-UV continuum.  Standard $\alpha$-discs with $\Mexp\approx1$ and $\Mdot/\Medddot\approx0.3$ cannot create $\alpha_{1450-4200\AA}=-0.5$.  A soft optical power-law corresponding to observations can be achieved by the standard $\alpha$-disc model only for large $\MBH$ or low accretion rates.\label{fig:M8_Mdot}}
\end{figure*}

We find that our wind models are able to soften the optical-UV SED (decrease the value of $\alpha$) creating a result more similar to observations of large samples of AGNs.  This can be seen by comparing the \cite{VandenBerk01} composite to our models in fig. \ref{fig:cat2_specs}.  Changes in the observed $\alpha$ can also be caused by a different combination of BH mass and accretion rates, in cases without wind.  We have therefore searched for standard $\alpha$-disc models which give similar SEDs to those of our modified model.  We allowed freedom of all relevant disc parameters ($\MBH$, $\Mdot$, $a$) and created SEDs with optical-UV power-laws $\alpha_{1450-4200\AA}\approx-0.5$.  Fig. \ref{fig:M8_Mdot} shows contours of $\alpha_{1450-4200\AA}=-0.5$ for a large range of $\Mexp$, $\Mdot$ and various values of $a$.  We were only able to model such soft optical and UV spectra by choosing $\Mexp\gg1$ or $\Mdot/\Medddot\ll0.1$ which both have the effect of softening the optical continuum.  We were unable to model soft enough optical-UV spectra with $\Mexp\approx1$ and $\Mdot/\Medddot\approx0.1-0.3$.  This has important implications, since such soft continua are observed in objects where the BH mass is observed to be around $\Mexp=1$.

\subsection{Effects of Wind on the Estimation of Black Hole Growth Rate}
\label{subsec:MBHdot}

\begin{table*}
\center
\caption{Changes in factors effecting estimation of $\MBHdot$ for wind emitting $\alpha$-disc models.  $\Lo$ is defined in eqn. \ref{eq:Lo=Mdot}; $b(5100\AA)$ is the bolometric correction calculated according to $b(\lambda)=L/\lambda\Llambda$; $b_M$ is the bolometric correction as calculated by the Marconi (2004) method at $5100$\AA; $b_M/b(5100\AA)$ is equal to $(b_M\lambda\Llambda)/L$; $b_M\lambda\Llambda/\Lo$ is the error in evaluation of $\MBHdot$ if only a measurement of $\lambda\Llambda$ is available and therefore may be written $\dot{M}_{BHestimated}/\dot{M}_{BHreal}$.}
\begin{tabular}{l|c|c|c|c|c|c}
\hline
\hline
Model       & $L/\Ledd$ & $L/\Lo$ & $b(5100\AA)$ & $b_M$ & $b_M/b(5100\AA)$ & $b_M\lambda\Llambda/\Lo$ \\
\hline
\hline

Model 1.1   & 0.1       & 1       & 8.95         & 7.41  & 0.83             & 0.83                     \\

Model 1.2   & 0.23      & 2.37    & 10.41        & 7.028 & 0.68             & 1.60                     \\

Model 1.3   & 0.14      & 1.42    & 7.24         & 7.10  & 0.98             & 1.40                     \\

Model 1.4   & 0.11      & 1.11    & 7.50         & 7.25  & 0.97             & 1.07                     \\

\hline

Model 1.5   & 0.01      & 1       & 4.69         & 8.61  & 1.83             & 1.83                     \\

Model 1.6   & 0.014     & 1.42    & 3.94         & 8.17  & 2.08             & 2.95                     \\

Model 1.7   & 0.029     & 2.90    & 3.60         & 7.61  & 2.11             & 6.13                     \\

Model 1.8   & 0.071     & 7.14    & 4.07         & 7.16  & 1.76             & 12.56                    \\

\hline
\hline

Model 2.1   & 0.1       & 1       & 16.50        & 7.80  & 0.47             & 0.47                     \\

Model 2.2   & 0.16      & 1.66    & 13.79        & 7.37  & 0.53             & 0.87                     \\

Model 2.3   & 0.12      & 1.20    & 11.71        & 7.46  & 0.64             & 0.76                     \\

Model 2.4   & 0.103     & 1.05    & 13.79        & 7.65  & 0.55             & 0.58                     \\

\hline

Model 2.5   & 0.01      & 1       & 8.41         & 9.17  & 1.091            & 1.091                    \\

Model 2.6   & 0.012     & 1.20    & 6.57         & 8.76  & 1.33             & 1.60                     \\

Model 2.7   & 0.019     & 1.89    & 4.88         & 8.12  & 1.66             & 3.14                     \\

Model 2.8   & 0.038     & 3.86    & 4.48         & 7.57  & 1.69             & 6.52                     \\

\hline

Model 3.3   & 0.13      & 1.35    & 13.71        & 7.49  & 0.55             & 0.73                     \\

Model 3.6   & 0.013     & 1.35    & 7.40         & 8.76  & 1.18             & 1.59                     \\

Model 3.7   & 0.025     & 2.55    & 6.84         & 8.15  & 1.19             & 3.043                    \\

Model 3.8   & 0.059     & 6.00    & 7.43         & 7.61  & 1.02             & 6.15                     \\

\hline
\hline
\end{tabular}
\label{table:MBHdot}
\end{table*}

As discussed above, for a wind ejecting disc, there is no simple way of calculating $\MBHdot$ by measuring $L$.  Since $\MBHdot$ depends only on spin and $\Mindot$, and since models in each category have the same $\Mindot$, all those with equal spin have the same BH growth rate.  However the total $L$ and the spectral shape of each can greatly vary.  Table \ref{table:MBHdot} shows a summary of factors influencing the correlation between $\MBHdot$ and $L$.  $L/\Lo$ is the ratio of the luminosity of the model to that of a standard $\alpha$-disc model with $\Mdot=\Mindot$ ($\Lo$ is defined in eqn. \ref{eq:Lo=Mdot}).  For models with the same $\Mindot$ and $\Moutdot$ but different $\rhalf$, a disc which loses most of its mass at smaller radii (models 1.2 and 2.2) result in larger $L/\Lo$ than a disc which loses most mass at larger radii (model 1.4 and 2.4), because the luminosity of the latter case is smaller.  Obviously, $L/\Lo$ can reach extremely large values when $\Moutdot/\Mindot$ is large.  For our most extreme models, $L/\Lo\approx7$.

The standard way to estimate $L$ from observations is to use a wavelength dependent bolometric correction factor, $b(\lambda)=L/\lambda\Llambda$ at a chosen wavelength.  For example $b(5100\AA)$ is used to estimate $L$ in low redshift AGNs.  Our calculated values of $b(5100\AA)$ and the standard bolometric correction calculated by the \cite{Marconi04} method for $5100$\AA~ ($b_M$) are given in table \ref{table:MBHdot}.  We note that the \cite{Marconi04} method includes contributions to $L$ from both the optical-UV and x-ray continua, whereas we do not include the x-ray continuum in our models and therefore calculate $L$ without the x-ray contribution.  The same method, when used for $L(x-ray)$, integrated over the x-ray continuum, results in $L(x-ray)/L$ in the range $\sim1-30\%$ depending on $L$ and on the energy range used for integration.  Therefore, the fact that we do not take this luminosity contribution into account may lead to differences in bolometric correction factor of up to $\sim20\%$ for extreme cases and has a minor effect on the bolometric correction calculations for most of our models.  The ratio $b_M/b(5100\AA)$ depends on the spectral shape of the AGN.  Values of this ratio close to unity imply accurate estimates of the disc luminosity even in the presence of wind.  Multiplying $L/\Lo$ by $b_M/b(5100\AA)$ (3rd and 6th columns of the table) gives the ratio $(b_M\lambda\Llambda)/\Lo$ which is the error in evaluation of $\MBHdot$ if only a measurement of $\lambda\Llambda$ is available.  These values are given in the last column of the table.

We see in table \ref{table:MBHdot} that for models with $\Moutdot/\Mindot=3$, $L/\Lo$ is in the range 1 - 2.37 for models with $a=0$, and in the range 1 - 1.66 for models with $a=0.9$.  The minimum values correspond to large $\rhalf$ and the maximum values to small $\rhalf$.  The values of $b_M/b(5100\AA)$ are in the range between 0.47 - 2.11 for all models, indicating a possible deviation of factor $\sim2$ if wind is not taken into account.  Values of $b_M\lambda\Llambda/\Lo$ for models with $\Moutdot/\Mindot=3$ range between 1.07 - 2.95 for models with $a=0$, and between 0.58 - 1.6 for models with $a=0.9$.  In general, higher spin values give lower ratios of $L/\Lo$, $b_M/b(5100\AA)$ and their product, corresponding to smaller errors in estimation of $\MBHdot$ from $L$ assuming a standard (constant $\Mdot$) $\alpha$-disc.  Models with larger ratios of $\Moutdot/\Mindot$ may have even higher errors in estimation of $\MBHdot$.  For our most extreme cases, with $\Moutdot/\Mindot=30$, we get $b_M\lambda\Llambda/\Lo\approx12.5$.

For a standard $\alpha$-disc model, when $\Mexp$ is known, one may estimate $\Mdot$ using the method of \cite{DavisLaor11}, utilizing the fact that $\Lnu$ is insensitive to BH spin at long wavelengths.  For a wind ejecting model, use of this method may cause significant errors in estimation of the accretion rate at the ISCO and consequently the BH growth rate.  We have calculated the ratio of $\Mdot$ calculated according to the \cite{DavisLaor11} method to the real $\Mindot$ for all our models and find significant deviations for models with $\Moutdot/\Mindot>1$.

\begin{figure*}
\begin{minipage}{\textwidth}
\center
\begin{tabular}{cc}
\includegraphics[width=0.5\textwidth]{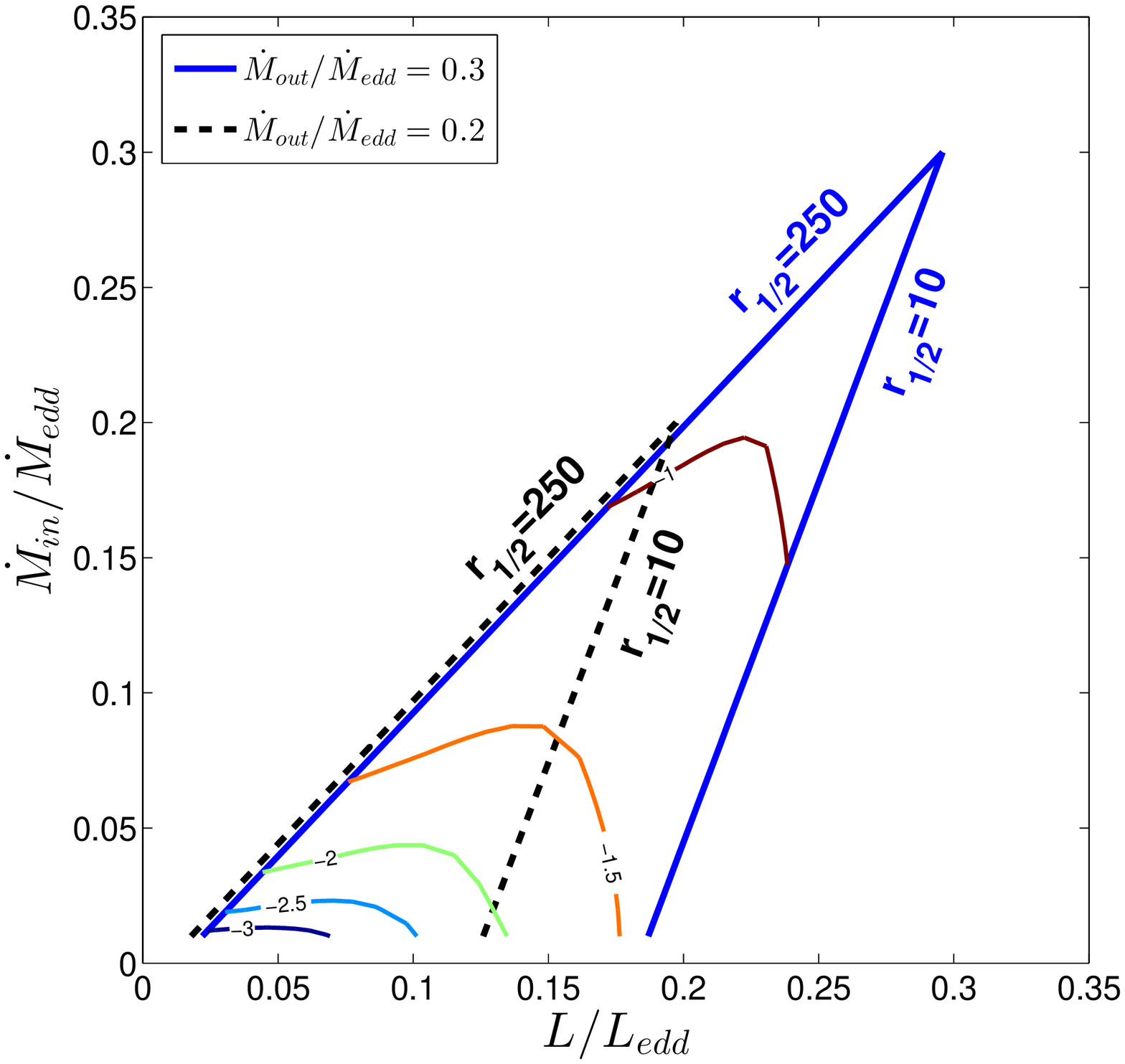} & \includegraphics[width=0.5\textwidth]{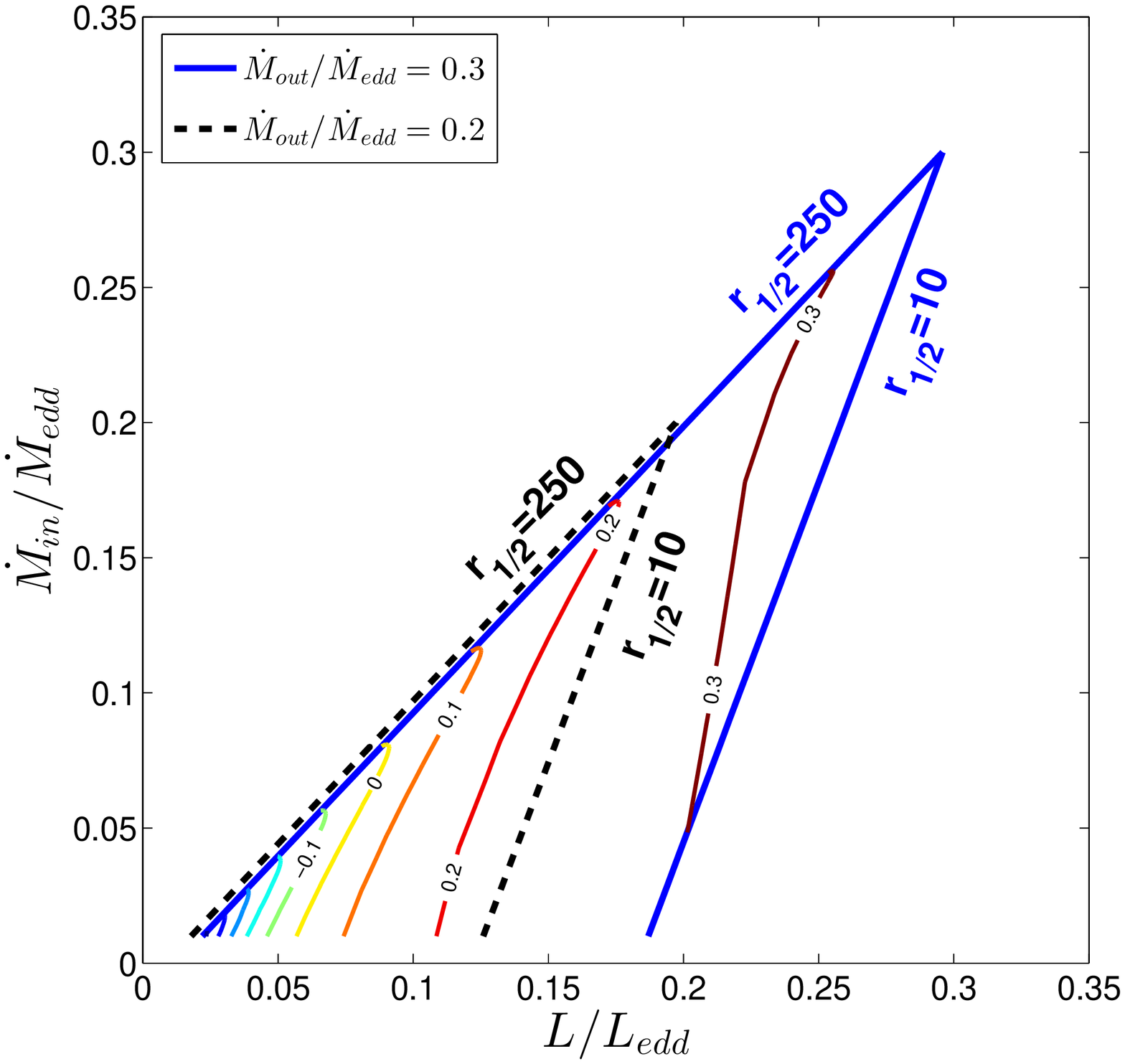}
\end{tabular}
\caption{Sensitivity of UV and optical power-laws to $\Mindot$ and $L$.  Shown in blue solid lines are borders of possible values for $\Mindot/\Medddot$ as a function of $L/Ledd$ for AGNs with $\Mexp=1$, $\Moutdot/\Medddot=0.3$ and $a=0$ and all possible wind profiles (i.e. all values of $\rhalf$).  Each coordinate within these boundaries corresponds to a possible model with a certain $\Mindot$ and $L$ and varying $\Mrdot$ profiles.  Shown in dashed black lines are the same borders for AGNs with $\Mexp=1$, $\Moutdot/\Medddot=0.2$ and $a=0$.  Colored contours are shown of constant power-law values for two wavelength ranges (UV and optical) from table \ref{table:slopes} - {\it Left}: $\alpha_{456-912\AA}$ and {\it Right}: $\alpha_{4200-5100\AA}$.  These contours are shown for the $\Moutdot/\Medddot=0.3$ models only, they differ from those of the $\Moutdot/\Medddot=0.2$ models but retain the same general trends.  $\alpha_{456-912\AA}$ (UV) is far more sensitive than $\alpha_{4200-5100\AA}$ (optical) to changes in $\Mindot$, therefore, knowledge of $\alpha_{456-912\AA}$ best constrains $\Mindot$ for a given $L$.\label{fig:L_Mindot_slopes}}
\end{minipage}
\end{figure*}

We have calculated the values of $L/\Ledd$ for models with $\Mexp=1$, $a=0$, $\Moutdot/\Medddot=0.3$ and all wind profiles.  This results in a continuous range of values for $\Mindot/\Medddot=0.01-0.3$ and $\rhalf=10-250$.  The blue solid lines in fig. \ref{fig:L_Mindot_slopes} show $\Mindot/\Medddot$ as a function of $L/\Ledd$ for two values of $\rhalf$, $10$ and $250$.  Each coordinate within these boundaries corresponds to a possible model with a certain $\Mindot$ and $L$.  For a given $L/\Ledd$, there is a broad range of allowed values of $\Mindot/\Medddot$ for various wind profiles.  The diagram shows that there is a value of $L/\Ledd$ for which there is a maximum range of $\Mindot/\Medddot$, corresponding to the luminosity for which we have maximum uncertainty in the accretion rate profile (i.e. in the form of $\Mrdot$).  We have calculated the values of $\alpha$ for the four wavelength ranges shown in table \ref{table:slopes} for all models within the parameter space confined by the blue lines of fig. \ref{fig:L_Mindot_slopes} (i.e. those with $\Mexp=1$, $a=0$ and $\Moutdot/\Medddot=0.3$).  The figure shows contours of $\alpha_{456-912\AA}$ (left) and $\alpha_{4200-5100\AA}$ (right).  $\alpha$ is most sensitive to changes in $\Mindot$ in the UV part of the spectrum.  We show this result only for $\Moutdot/\Medddot=0.3$, but it holds true for all values of $\Moutdot$, with the actual shapes of the contours varying for different values of $\Moutdot$.  In each panel of fig. \ref{fig:L_Mindot_slopes} we show in dashed black lines the borders of $L/\Ledd$ vs $\Mindot/\Medddot$ for AGNs with $\Mexp=1$ and $\Moutdot/\Medddot=0.2$.  If measurements of $L$ and of $\alpha$ in the UV are available for an AGN, one can constrain the value of $\MBHdot$ by utilizing the sensitivity of $\alpha$ in the UV to $\Mindot$ and taking into consideration the unknown parameters $\Moutdot$ and $a$.

\section{Discussion}
\label{sec:discussion}

The wind ejecting disc model presented in this work is motivated by two observed phenomena: discrepancies between observed optical-UV SEDs of AGN to those predicated by standard $\alpha$-disc models \cite[as found by][]{Davis07}, and evidence for high velocity winds which may be ejected from the inner AD.  The models calculated here are not intended to represent real disc winds.  The properties of such winds may depend, among other factors, on disc turbulence, the rate and direction of mass inflow, disc geometry and magnetic fields, all not very well understood.  Our limited scope calculations are meant to study the influence of general properties of winds and their radius dependent mass loss, on the fundamental disc properties such as total luminosity, SED shape and BH growth rate.  Obviously, the calculations do not cover all such possibilities.  Furthermore, we stress that in order to preserve energy conservation within our model, there are a number of possibilities regarding the kinetic energy within the ejected wind.  As stated above, such possibilities include that the wind remains in the AD vicinity and therefore does not require excess kinetic energy for escape from the BH potential.  Another possibility is that the wind's excess kinetic energy is liberated from photons which are not along our line of sight and therefore does not affect the observed SED.  Including all such considerations, there are two main results of the new calculations.  The first is the understanding that disc winds are able to considerably soften the optical-UV SED of accretion discs.  Second, we have shown that the estimation of $\MBHdot$ cannot rely solely on measurment of luminosity, as $L/\Ledd$ can differ substantially from $\Mindot/\Medddot$.

The SED of a wind ejecting disc is influenced by the fact that removal of hot accreting gas from the inner regions of the disc removes energy mainly from the UV part of the SED and has less effect on the optical and NIR continua.  The extent of this energy removal depends on $\Mindot$, $\Moutdot$ and on the shape of the wind.  $\Moutdot$ represents the accretion rate over most of the disc and therefore adjusts the peak wavelength of the SED and contributes radiation mainly at low frequencies, with a lesser contribution to the optical-UV continuum which depends on $\Mindot$ and $\rhalf$.  Furthermore, the contribution to the long wavelength continuum is almost frequency independent whereas the contribution to the UV is frequency dependent.  $\Mindot$ governs the shape of the UV continuum such that low values of $\Mindot$ compared to $\Moutdot$ soften the UV SED significantly.  $\rhalf$ governs the relative importance of accretion rates at the inner and outer radii of the disc.  Specifically, a large value of $\rhalf$ causes energy to be removed at larger radii, influencing the SED over a large range of wavelengths from NIR to UV.  A small $\rhalf$ causes an almost constant accretion rate throughout almost the entire disc and a sharp fall of accretion at small radii, influencing mainly the UV SED.  These factors together, determine the spectral shape of a wind ejecting disc and are evident in figures \ref{fig:cat1_specs} - \ref{fig:cat2_specs}.  We find that we can model the observed $\alpha_{1450-4200\AA}\approx-0.5$ for $\Mexp=1$ BHs and a range of $\Mindot$, $\Moutdot$, $a$ and $\rhalf$.  Such soft spectra cannot be modelled by standard $\alpha$-discs except for very large BH mass or very low accretion rates.  In particular, some of our models (1.8 and 2.7) fit relatively well with the \cite{VandenBerk01} composite shown in fig. \ref{fig:cat2_specs}.  We note that these models have extremely high ratios of $\Moutdot/\Mindot$ and that models 1.5 and 2.5 (which are standard $\alpha$-disc models with the same $\Mindot$) already have relatively soft optical-UV spectra.

Since eqn. \ref{eq:L=Mdot} is not valid for wind ejecting discs, $\MBHdot$ cannot be evaluated solely by measurement of $L$.  As fig. \ref{fig:L_Mindot_slopes} shows, there is a range of values, $\Mindot/\Medddot$, for a given $L/\Ledd$.  We have found that estimating $L$ by using a certain $\lambda\Llambda$ and a bolometric correction may result in inaccuracy by a factor $\sim2$.  Furthermore, we find that the method suggested by \cite{DavisLaor11} for estimation of $\Mdot$ by measuring $\Lnu$ at optical wavelengths is not consistent with our wind ejecting models and may cause significant discrepencies in evaluation of $\MBHdot$.  Together with knowledge of $L$, information on the spectral shape of the AGN can further constrain $\MBHdot$.  In fig. \ref{fig:L_Mindot_slopes}, comparing the left and right panels, one sees that the slope, $\alpha$, in some spectral ranges is more sensitive to $\Mindot$ than in other spectral ranges.  For example, $\alpha$ at UV wavelengths is far more sensitive to changes in accretion rate at the ISCO, than at NIR wavelengths.  This is true for all values of $\Mexp$, $a$ and $\Moutdot$.  Therefore, for known $\Mexp$ and $a$ values, measurement of $\alpha_{456-912\AA}$ or $\alpha_{912-1450\AA}$ can remove much of the uncertainty in $\MBHdot$ caused by different wind profiles and leaves only the uncertainty caused by the unknown value of $\Moutdot$.

Finally, this work is not intended to explore the physics of real AGN winds or their influence on the disc geometry and physical state.  As explained above, there are various types of winds (radiation driven, thermally driven, MHD) and a large range of factors which may influence the outflows.  Our work shows that such winds affect the disc luminosity, the SED and the BH growth rate, under very general assumptions about the actual wind profile.  The understanding of the effects of disc winds on the SED and on our ability to measure $L$ allows one to constrain the possible errors made in estimation of BH growth rate.

\section{Acknowledgements}
\label{sec:Acknowledgements}

We thank Ari Laor for a fruitful discussion regarding our models.  We are grateful to Benny Trakhtenbrot and Rivay Mor for their advice and contributions to this publication.  Funding for this work has been provided by the Israel Science
Foundation grant 364/07.


\bibliographystyle{mn2e}
\bibliography{bibliography}


\end{document}